# Possible valence-bond condensation in the frustrated cluster magnet LiZn$_2$Mo$_3$O$_8$

J.P. Sheckelton, J.R. Neilson, D.G. Soltan, and T.M. McQueen

Department of Chemistry, Department of Physics and Astronomy, and the Institute for Quantum Matter, The Johns Hopkins University, Baltimore, MD 21218

The emergence of complex electronic behaviour from simple ingredients has resulted in the discovery of numerous states of matter. Many examples are found in systems exhibiting geometric magnetic frustration, which prevents simultaneous satisfaction of all magnetic interactions. This frustration gives rise to complex magnetic properties such as chiral spin structures[1,2,3] orbitally-driven magnetism[4], spin-ice behavior[5] exhibiting Dirac strings with magnetic monopoles[6], valence bond solids[7,8], and spin liquids[9,10]. Here we report the synthesis and characterization of LiZn$_2$Mo$_3$O$_8$, a geometrically frustrated antiferromagnet in which the magnetic moments are localized on small transition metal clusters rather than individual ions[11,12,13]. By doing so, first order Jahn-Teller instabilities and orbital ordering are prevented, allowing the strongly interacting magnetic clusters in LiZn$_2$Mo$_3$O$_8$ to probably give rise to an exotic condensed valence-bond ground state reminiscent of the proposed resonating valence bond state[14,15]. Our results also link magnetism on clusters to geometric magnetic frustration in extended **solids, demonstrating a new approach for unparalleled chemical control and tunability in the search for collective, emergent electronic states of matter[16,17].**

Numerous materials possess a geometrically frustrated arrangement of magnetic atoms; such materials have their magnetic moments arranged on frustrated topologies such as triangular lattices[18,19], kagomé lattices[20,21], hyper-kagomé lattices[22], and edge sharing tetrahedra[23,24]. A rich diversity of properties result depending on the magnitude of the per-site spin and orbital occupancies. Yet, the presence of local structural distortions[25,26], and a degree of site mixing between non-magnetic and magnetic layers[27] are still key limiters in the quest for new quantum states of matter[28]. Here we show that these problems can be overcome through the use of clusters in which a magnetic, unpaired electron is delocalized over a small number of transition metal atoms, rather than individual magnetic ions, by demonstrating that the $S = ½$ cluster magnet $LiZn_2Mo_3O_8$ is geometrically frustrated and likely possesses a condensed valence bond ground state.

$LiZn_2Mo_3O_8$ is built of discrete $Mo_3O_{13}$ cluster units (figure 1a), in which all Mo atoms are on equivalent crystallographic sites[29]. The average formal oxidation state of Mo is +3.67. Each cluster has seven valence electrons. By a simple electron count, supported by molecular orbital calculations (figure 1b), six of these electrons localize into Mo-Mo bonds holding the cluster together. The seventh electron remains unpaired in a totally symmetric (A1 irreducible representation) molecular orbital with equal contributions from all three Mo atoms. The result is one $S = ½$ magnetic moment on each $Mo_3O_{13}$ cluster. This cluster can then replace an atom as the basic building block of a geometrically frustrated magnetic system, when appropriately arranged. There have been previous reports of systems built of magnetic clusters[30]. In those cases, the unpaired, magnetic electrons within a cluster are still (just like the non-cluster cases) localized on individual atoms. By contrast, in LiZn$_2$Mo$_3$O$_8$, the magnetism arises from a collective contribution of all three atoms in the Mo$_3$O$_{13}$ cluster. This gives rise to a $S = \frac{1}{2}$ moment delocalized over three Mo atoms. This delocalized nature of the moment contributes to the stability of the system and renders the structure impervious to first order Jahn-Teller distortions.

These Mo$_3$O$_{13}$ clusters in LiZn$_2$Mo$_3$O$_8$ connect at corners, making a two-dimensional Mo$_3$O$_8$ layer (figure 1c). The 2.6 Å Mo-Mo distance within each cluster is substantially shorter than between clusters (3.2 Å) reflecting strong metal-metal bonding within each cluster. These Mo$_3$O$_8$ layers are separated by non-magnetic Li/Zn ions (figure 1d) to form the full structure with $R\bar{3}m$ symmetry. Consequently, LiZn$_2$Mo$_3$O$_8$ contains two-dimensional layers in which $S = \frac{1}{2}$ Mo$_3$O$_{13}$ clusters are arranged on the geometrically frustrated triangular lattice.

The temperature dependence of the magnetic susceptibility of LiZn$_2$Mo$_3$O$_8$ shows unusual and unexpected behavior. At temperatures above 96 K, the inverse magnetic susceptibility (figure 2a) is well described by the Curie-Weiss law for paramagnetic spins. A Weiss temperature of $\theta = -220$ K indicates a net mean-field antiferromagnetic interaction between unpaired spins on Mo$_3$O$_{13}$ clusters. A Curie constant of $C = 0.24$ emu·K·Oe$^{-1}$·mol $f.u.^{-1}$ ($p_{eff} = 1.39$) is reduced from the ideal 0.375 value for a free $S = \frac{1}{2}$ moment. This may be due to a number of possibilities (see SI), but the most likely is a partial unquenched orbital contribution to the moment, due to spin-orbit coupling. It is not due to the formation of a metallic state: resistivity data (figure S6) shows that LiZn$_2$Mo$_3$O$_8$ is electrically insulating at all accessible temperatures. Furthermore, the molecular calculations predict an on-site (cluster) Hubbard $U$ of ~1.2 eV, which, depending on bandwidth, could open a gap and explain the insulating behavior. Together, these data imply that the one unpaired electron per cluster in LiZn$_2$Mo$_3$O$_8$ behaves as a localized effective $S = \frac{1}{2}$ magnetic system (with a partial unquenched orbital contribution just like $Co^{2+}$ or $Cu^{2+}$), and that the magnetic interactions between clusters are strong and antiferromagnetic.

A change in the slope of the inverse magnetic susceptibility as a function of temperature occurs around $T = 96$ K, with a second linear region of $\chi(T)$ present below this transition (or crossover). A fit to the linear region from $T = 2$ K to $T = 96$ K gives a Weiss temperature of $\theta = -14$ K and a Curie constant of $C = 0.08$. This Curie constant is one-third that of the high temperature value, indicating that two-thirds of the spins contribute negligibly to the magnetic susceptibility below $T = 96$ K as the Curie constant scales with the number of moments.

Neutron powder diffraction experiments at $T = 12$ K indicate that long-range magnetic order does not develop below the $T \approx 96$ K transition (figure S1). Instead, our results are consistent with two-thirds of the effective spins condensing into magnetic singlets. Although our data are not sufficient to unambiguously determine whether these singlets are static, making a valence-bond solid, or dynamic, making a resonating valence-bond state, neutron powder diffraction data suggest that the singlets are indeed dynamic: at $T = 12$ K, $LiZn_2Mo_3O_8$ maintains the trigonal $R\bar{3}m$ symmetry that exists at $T = 300$ K. In most cases, static singlets form a valence bond network and distort the lattice to a lower symmetry. Unambiguous determination of the ground state warrants further study, but the ground state of $LiZn_2Mo_3O_8$ is unusual and reflective of the strong geometric magnetic frustration.

Changes in the experimentally measured heat capacity further elucidate the unusual electronic behavior in $LiZn_2Mo_3O_8$ (figure 2b). $LiZn_2Mo_3O_8$ does not undergo a transition to long range magnetic order above $T = 0.1$ K: there is no sharp lambda transition of the heat capacity as a function of temperature. Instead there is only an upturn in the specific heat capacity data below $T = 1$ K. Applied magnetic fields of $\mu_oH = 1$ T and $\mu_oH = 9$ T (figure 2b inset) radically modulate the behavior of the low temperature data. Such large changes from small magnetic fields are surprising given the large Weiss temperature and are likely a result of magnetic frustration in the system. Geometric frustration prevents the formation of long range order and results in low-lying magnetic excitations perturbed by an applied field. Simple models, such as a multilevel Schottky anomaly, do not adequately describe the low temperature data (see SI); further studies are needed to examine and understand the behavior in detail.

The magnetic entropy change of $LiZn_2Mo_3O_8$, accounting for the extra lattice contribution from lithium compared to $Zn_2Mo_3O_8$ (figure 2c, figure S2) also indicates the condensation of two-thirds of the available spins. The total expected magnetic entropy change for $S = ½$ system is $R \cdot \ln(2)$ (= 5.76 $J \cdot K^{-1} \cdot mol\, f.u.^{-1}$), compared to the experimental value of 8(3) $J \cdot K^{-1} \cdot mol\, f.u.^{-1}$ from $T = 0.1$ to $T = 400$ K. On cooling from $T = 400$ K, we observe a gradual and continuous loss of entropy, approximately two-thirds of the expected $S = ½$ value from $T = 400$ K to $T = 100$ K. Critically, the change in the linear regions of magnetic susceptibility is not accompanied by a sharp transition in the entropy, supporting the claim that these spins condense into singlets, rather than adopt long range magnetic order. Furthermore, the difference in entropy between $T = 0.1$ K and $T = 100$ K is approximately $\frac{1}{3} R \cdot \ln(2)$, consistent with freezing out of the remaining one-third of spins that did not condense into singlets at $T = 96$ K.

The resulting magnetic phase diagram of $LiZn_2Mo_3O_8$ is shown in figure 2d. Near room temperature, the system is paramagnetic and the spins thermally randomize. Cooling below the condensation temperature ($T \sim 96$ K), two-thirds of the spins form a condensed valence bond state. The remaining one-third spins are still paramagnetic and interacting antiferromagnetically until lower temperatures, at which point they lose entropy in a yet-to-be determined manner.

These results suggest that LiZn$_2$Mo$_3$O$_8$ exhibits geometric magnetic frustration between $S$ = ½ magnetic clusters and two-thirds of the spins condense into singlets below approximately $T$ = 96 K. Therefore LiZn$_2$Mo$_3$O$_8$ is a candidate for a resonating valence-bond state, as there is no evidence for static singlets. More generally, our results show how an extended lattice of magnetic clusters, in place of magnetic ions, produces exotic physics while providing numerous advantages in the design and control of magnetically frustrated materials. This approach opens a new chemical frontier in the search for emergent phenomena in frustrated systems.

# Acknowledgements

This research is supported by the U.S. Department of Energy, Office of Basic Energy Sciences, Division of Materials Sciences and Engineering under Award DE-FG02-08ER46544. Use of the Spallation Neutron Source was supported by the Division of Scientific User Facilities, Office of Basic Energy Sciences, US Department of Energy, under contract DE-AC05-00OR22725 with UT-Battelle, LLC. J.P.S. acknowledges the assistance of J. Hodges in collecting and analyzing powder neutron data from POWGEN/SNS. TMM acknowledges useful discussions with O. Tchernyshyov and C. Broholm.

# Methods

Phase-pure $LiZn_2Mo_3O_8$ was synthesized from a mixture of Mo, ZnO, $Li_2MoO_4$, and $MoO_2$ (99+% purity) with an overall starting formula of $LiZn_2Mo_3O_8(Li_2Zn_2O_3)_{0.2}$. Mo was used as received. ZnO and $Li_2MoO_4$ were dried at $T = 160$ °C overnight. $MoO_2$ was purified by heating overnight under flowing 5% $H_2$/95% Ar. The mixtures were pressed into pellets, placed in alumina crucibles, and double-sealed in evacuated, fused silica tubes. The reaction vessel was heated to $T = 600$°C for 24 hours, ramped to $T = 1000$°C at $10$°C/hr, held for 12 hours, followed by a water quench. The sample was reground and heated again in the same manner. $Zn_2Mo_3O_8$ was synthesized in a similar manner, but with 3% excess ZnO, and a final temperature of $T = 1050$ °C.

Magnetization measurements, heat capacities, and resistivities were measured on a sintered pellet in a Quantum Design Physical Properties Measurement System (PPMS) using a dilution refrigerator for $T < 2$ K measurements. Heat capacities were measured three times at each temperature using the semi-adiabatic pulse technique, waiting for three time constants per measurement. Data were collected from $T = 0.05$ K to $T = 400$ K under magnetic fields of $\mu_oH = 0$ T, $\mu_oH = 1$ T and $\mu_oH = 9$ T. Magnetic susceptibilities were measured from $T = 1.8$ K to $T = 320$ K under a $\mu_oH = 1$ T field. Laboratory X-ray powder diffraction patterns were collected using Cu Kα radiation (1.5418 Å) on a Bruker D8 Focus diffractometer with a LynxEye detector. Powder neutron diffraction data sets at $T = 12$ K with $d$-spacing of 0.2760Å $\leq d \leq$ 3.0906Å and 1.6557Å $\leq d \leq$ 8.2415Å were collected at the Spallation Neutron Source POWGEN diffractometer (BL-11A) at the Oak Ridge National Laboratory and analyzed with the Rietveld method using GSAS with EXPGUI[31,32]. Molecular orbital calculations using density functional theory with the PBE0 functional at the UHF level of theory were performed with the GAMESS software package[33].

# Author Contributions

TMM supervised the project. JPS and DES prepared samples. JPS measured neutron diffraction patterns. JPS and TMM measured properties, and JPS, JRN, and TMM analyzed data and prepared the manuscript.

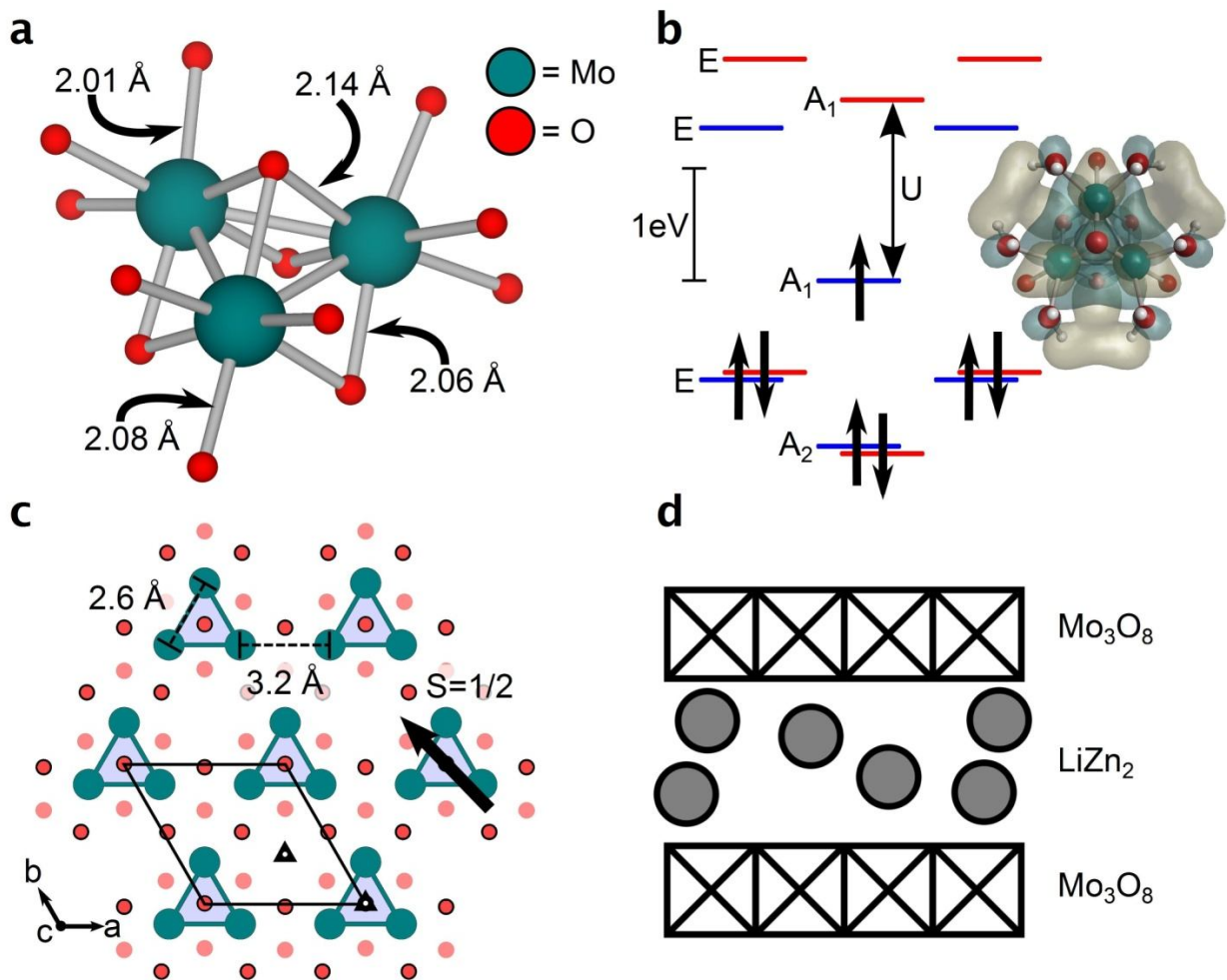

**Figure 1 | LiZn$_2$Mo$_3$O$_8$ structure. a,** A single Mo$_3$O$_{13}$ cluster shows the local coordination of each Mo atom. **b,** A spin polarized molecular orbital diagram for Mo$_3$O$_{13}$H$_{15}$ ($C_{3v}$). There is one unpaired electron per cluster, distributed over all Mo atoms, with a large energy gap to the next available state. The hybrid functional produces an estimate of the on-site repulsion energy, U ~ 1.2 eV. A$_1$, A$_2$ and E are the irreducible representation labels for each orbital level from the C$_{3v}$ point group. **c,** Top-down view of the Mo$_3$O$_8$ layer showing the triangular network formed by the Mo$_3$O$_{13}$ $S = ½$ clusters. **d,** A schematic representation of the magnetic Mo$_3$O$_8$ layers separated by LiZn$_2$ in LiZn$_2$Mo$_3$O$_8$.

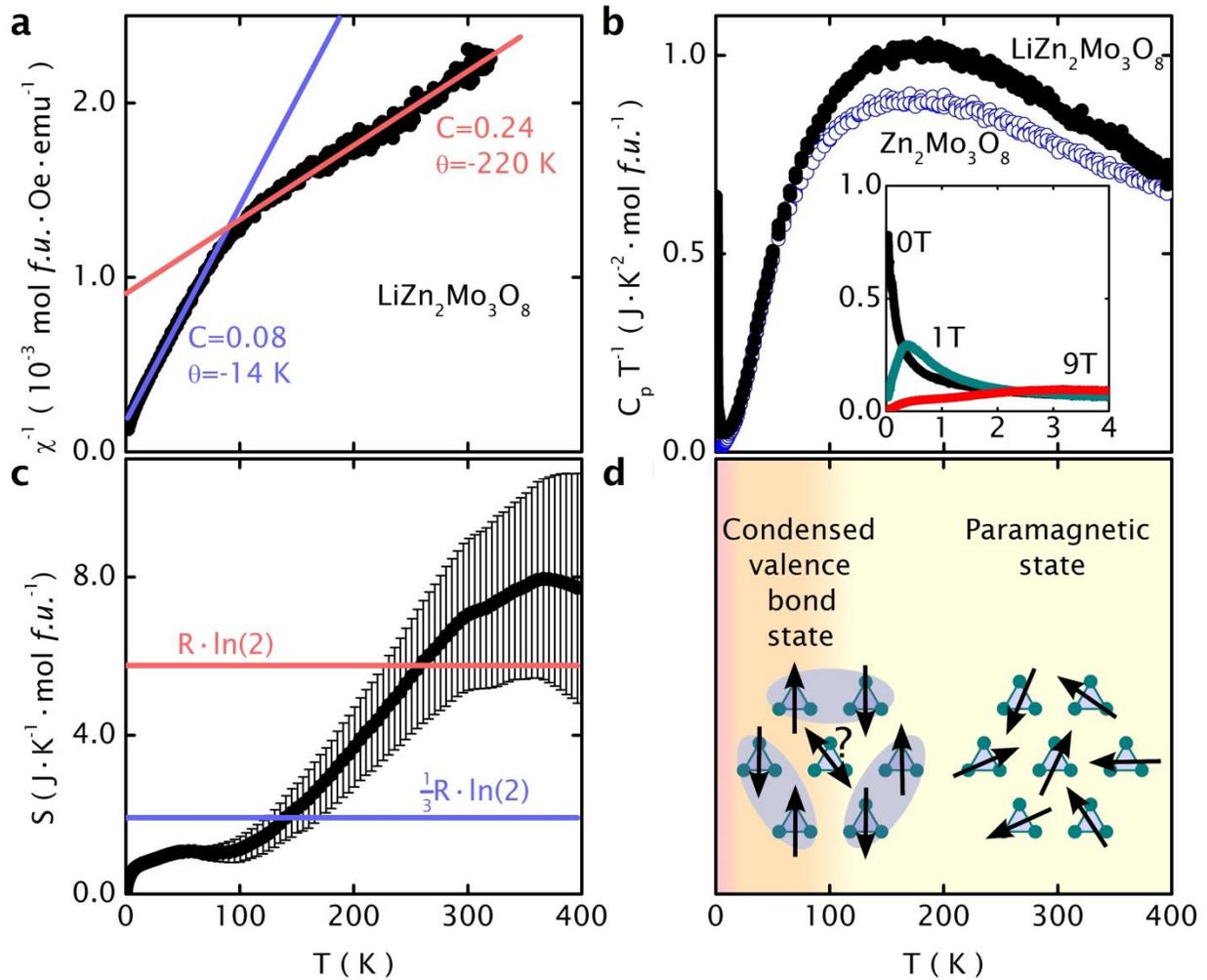

**Figure 2 | Physical properties of LiZn$_2$Mo$_3$O$_8$. a,** Inverse magnetic susceptibility as a function of temperature for LiZn$_2$Mo$_3$O$_8$. Curie-Weiss fits to the two distinct linear portions are shown. Two-thirds of the spins 'disappear' upon cooling below $T$ = 96 K. The Curie constant C is in units of emu·K·Oe$^{-1}$·mol $f.u.^{-1}$. **b,** Heat capacity divided by temperature as a function of temperature. The inset shows a strong magnetic field dependence of the low temperature specific heat. Data for non-magnetic Zn$_2$Mo$_3$O$_8$ is shown for comparison. **c,** Integrated entropy as a function of temperature. The lattice contribution was subtracted prior to integrating (see SI). Error bars are calculated using standard analysis of error techniques for the propagation of the uncertainty in each C$_p$ measurement through the numerical integration. This is given by

$$\delta S_N = \sum_{i=1}^{N} \frac{x_{i+1} - x_i}{2} \sqrt{(\delta y_{i+1})^2 + (\delta y_i)^2}$$, where the error bars are given by $\delta S_N$, and $\delta y_i$ is the uncertainty in the $C_p/T$ value of the $i^{th}$ point. **d,** Proposed magnetic phase diagram of $LiZn_2Mo_3O_8$. Below $T = 96$ K the spins enter a condensed valence bond state.

Supplementary information

# Possible valence bond condensation in the frustrated cluster magnet $LiZn_2Mo_3O_8$

J.P. Sheckelton, J.R. Neilson, D.G. Soltan, and T.M. McQueen

Department of Chemistry, Department of Physics and Astronomy, and the Institute for Quantum Matter, The Johns Hopkins University, Baltimore, MD 21218

email: mcqueen@jhu.edu

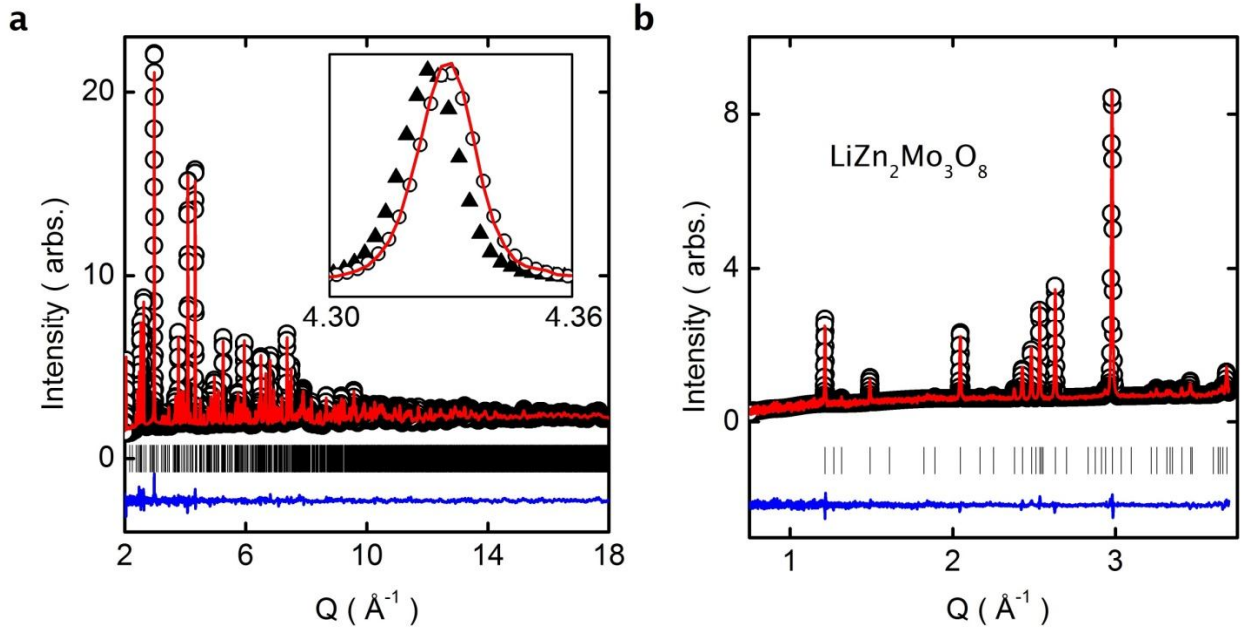

**Figure S1 | 12 K TOF neutron powder diffraction data for LiZn$_2$Mo$_3$O$_8$ from POWGEN**. **a,b.** Rietveld refinement of data show that LiZn$_2$Mo$_3$O$_8$ remains in the $R\bar{3}m$ spacegroup and maintains its trigonal symmetry in the condensed valence bond state. The inset in **a** shows the (220) peak at both $T = 12$ K (open circles) and at $T = 300$ K (triangles), which are sharp and show no indication of a distortion breaking trigonal symmetry. Attempts to fit the patterns to lower symmetry but trigonal models (R3m, R32, R3) did not result in better fits to the data. The patterns show no extra Bragg reflections or diffuse scattering due to magnetic ordering of any kind down to $T = 12$ K.

As a test, a magnetic phase was added to the refinement to asses if scattering from a magnetic phase would be visible. The magnetic form factor in this material is not known, and there are many possible magnetic orders. Thus to estimate our sensitivity to magnetic order, we used a 120° magnetic state with the Mo metal form factor. This resulted in an upper limit on the magnetic moment of 0.2(2) $\mu_B$ per Mo. Together with the absence of an anomaly in the heat capacity, these data suggest no magnetic ordering around $T = 96$ K.

**Heat capacity analysis.**

The magnetic contribution to the heat capacity was extracted by two methods. In the first method, the data for non-magnetic $Zn_2Mo_3O_8$ was scaled to account for the expected change in Debye temperature (compared to $LiZn_2Mo_3O_8$) as well as for the change in the number of atoms per formula unit (=14/13). Subtraction gives the estimated magnetic heat capacity shown in figure S3(a).

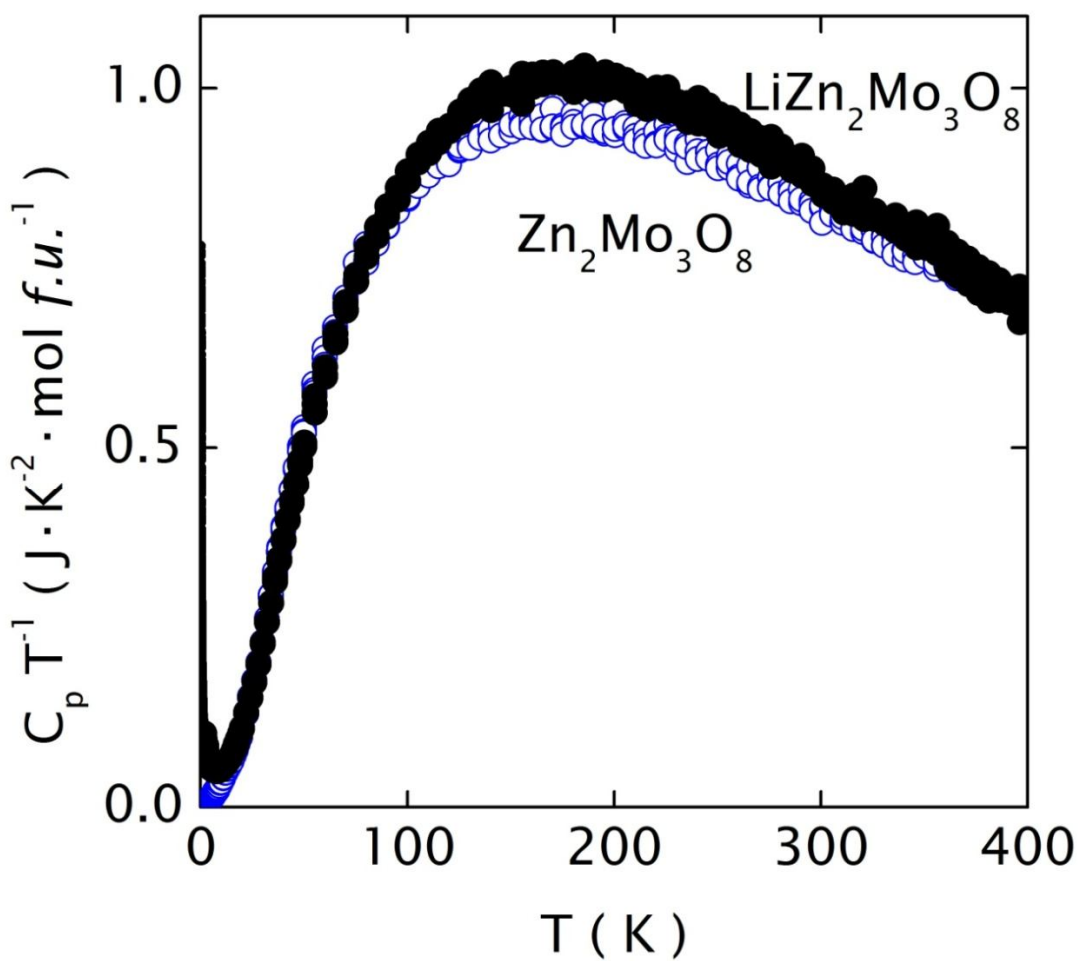

**Figure S2 | $LiZn_2Mo_3O_8$ and formula scaled $Zn_2Mo_3O_8$ $C_p \cdot T^{-1}$ datasets.**

In the second method, the non-magnetic $Zn_2Mo_3O_8$ data were not scaled for the change in the number of atoms per formula unit (paper Figure 2b). Instead, a smooth fit to the non-magnetic $Zn_2Mo_3O_8$ $C_p \cdot T^{-1}$ was directly subtracted, giving the data in figure S3(b), which includes both the magnetic contribution and the extra lattice contribution from the extra lithium atom per formula unit.

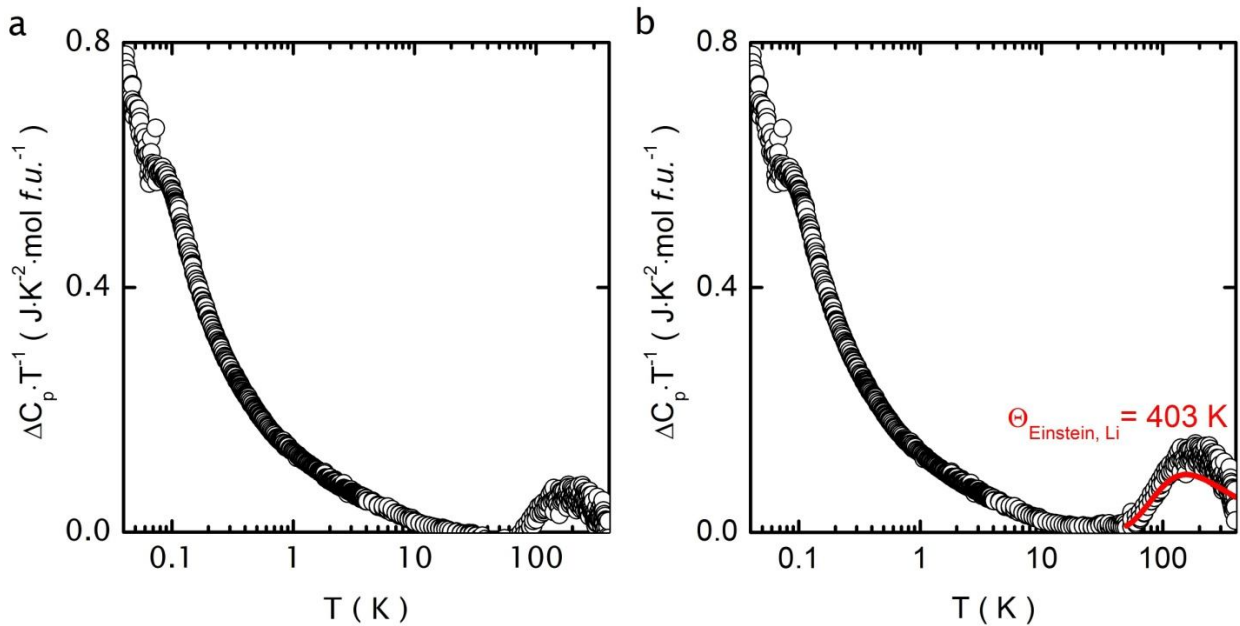

**Figure S3 | Estimated excess heat capacity in $LiZn_2Mo_3O_8$ computed by two methods. a,** In the first method, the data for non-magnetic $Zn_2Mo_3O_8$ was scaled to account for the expected change in Debye temperature (compared to $LiZn_2Mo_3O_8$) as well as for the change in the number of atoms per formula unit, leaving only an estimate of the magnetic entropy. Note the unphysical dip to negative heat capacity around $T = 50$ K. **b,** In the second method, the non-magnetic $Zn_2Mo_3O_8$ data were not scaled for the change in the number of atoms per formula unit, leaving contributions from both magnetism and the extra lattice contribution from the extra lithium atom per formula unit. The extra lattice entropy of Li can then be accounted for by fitting to an

Einstein (or Debye) oscillator mode (fit shown in red).

Both methods give similar insights into the magnetic behavior for $LiZn_2Mo_3O_8$. Method two gives a larger feature at $T \geq 100$ K, which must (at least partly) the freezing out of the extra vibrational modes from Li in $LiZn_2Mo_3O_8$. Figure S3(b) shows a fit of this feature to an Einstein oscillator mode, with an Einstein temperature $\Theta = 403$ K (a Debye mode fits equally well), which was subtracted to leave the magnetic contribution. In both cases, the magnetic entropy was then extracted by computing $S(T) = \int_0^T \frac{C}{T} dT$. A comparison of the two are shown in figure S4.

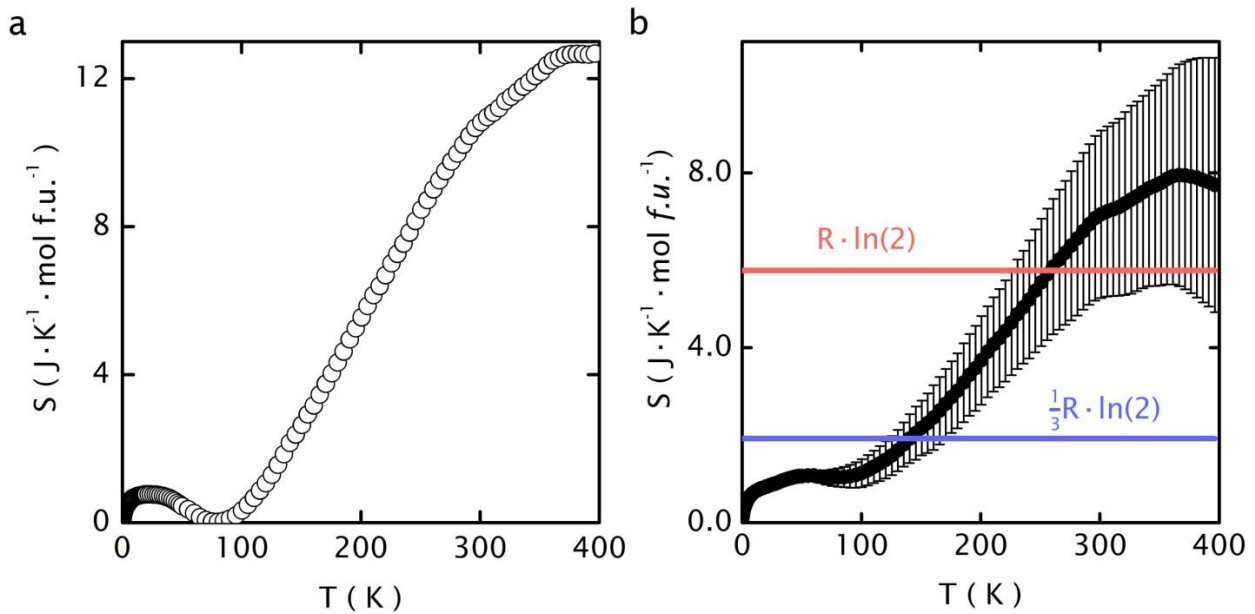

**Figure S4 | Estimated magnetic entropy by two methods**. **a,** Integrated $C_p \cdot T^{-1}$ data from method one. Although the dip in entropy around $T = 50$ K is unphysical, the general result, of two distinct regions of entropy loss - below and above $T \approx 100$ K – is the same as that found by method two. **b,** Integrated $C_p \cdot T^{-1}$ from method two. This data is the same as paper figure 2c, and also shows two distinct regions of entropy loss.

In both cases, the magnetic entropy contribution contains two discrete regions of entropy gain/loss, below and above $T = 100$ K. We note that the validity of fitting $C_p$ to an Einstein (or Debye) oscillator model for the lattice contribution at high temperature (rather than $C_v$), is justified. We estimate that $C_v$ and $C_p$ are the same to within approx. 1% up to $T = 400$ K: the relation $C_P - C_v = AC_p^2 T$ can be used to estimate $C_v$, where $A = \beta^2 V / C_p^2 \kappa_T$, $\beta = V^{-1}\left(\frac{\partial V}{\partial T}\right)_P$ is the volume expansion coefficient and $\kappa_T = -V^{-1}\left(\frac{\partial V}{\partial P}\right)_P$ is the isothermal compressibility[34]. LeBail refinements of the neutron diffraction patterns from $T=300$ K to $T=12$ K were used to estimate $\beta$ ($\approx 10^{-5}$ K$^{-1}$). The isothermal compressibility was estimated using a literature value for another closest packed oxide (MgO $\kappa_T = 7.2$ Pa$^{-1}$; virtually all oxides have isothermal compressibilities within one order of magnitude of this value[35]).

In both methods, there is a low temperature feature in the magnetic $C_p \cdot T^{-1}$ data. Figure S5 shows an analysis indicating that the feature cannot be adequately explained by simple Schottky anomaly models. Two or three level Schottky anomalies do not account for excess heat capacity in the data, even when other extra terms that might be present are also included.

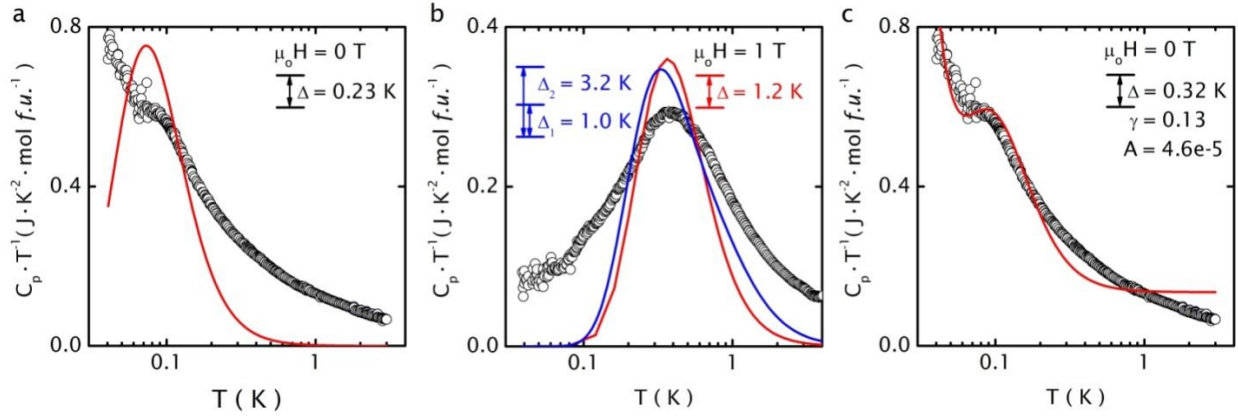

**Figure S5 | Low temperature $C_p \cdot T^{-1}$ Schottky anomaly analysis**. **a,** $\mu_o H = 0$ T data fit to a two level anomaly and **b,** $\mu_o H = 1$ T fit for both a two (red) and three (blue) level Schottky anomaly. **c,** A fit to the $\mu_o H = 0$ T data allowing for a $\gamma T$ electronic contribution ($\gamma$ is in units of $J \cdot K^{-2} \cdot mol\ f.u.^{-1}$) and $AT^{-2}$ nuclear contribution (A is in units of $J \cdot K \cdot mol\ f.u.^{-1}$). Despite allowing for these contributions (a $\gamma T$ term is unphysical considering the material is an insulator, see figure S6, although such a contribution could arise from spin-liquid behavior) the fit still does not adequately describe the data. In all cases, changing the degeneracy of the anomaly levels ($g_0/g_1$) does not resolve the fit, only changes the width by a small amount. All figures show datasets from both methods of subtracting the lattice contribution of $Zn_2Mo_3O_8$, but at these temperatures, the difference is negligible.

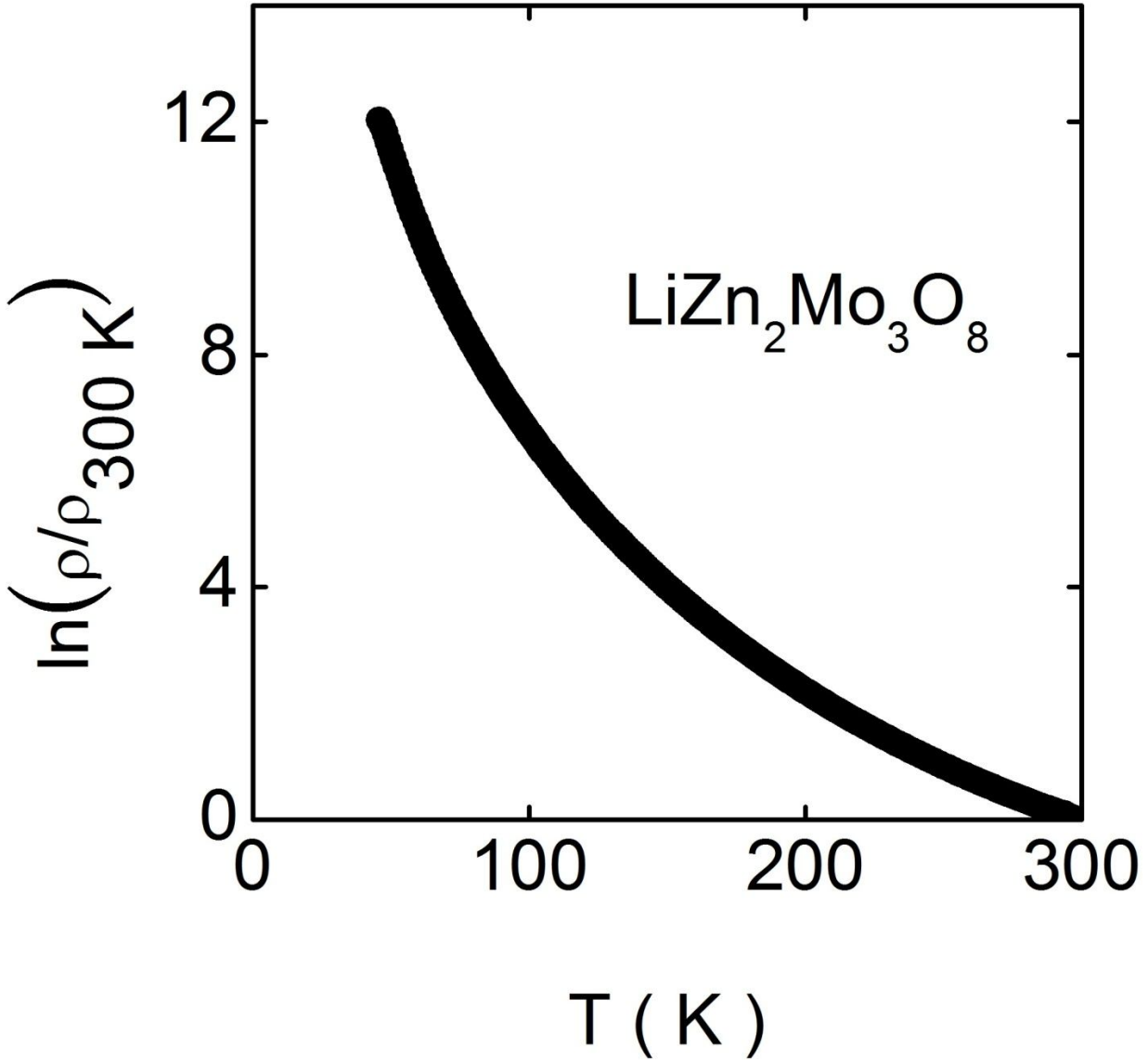

**Figure S6 | LiZn$_2$Mo$_3$O$_8$ resistivity.** Resistivity as a function of temperature data measured on a sintered pellet of LiZn$_2$Mo$_3$O$_8$. Measurements taken down to $T$ = 50 K (where the voltmeter saturated) show that LiZn$_2$Mo$_3$O$_8$ is an electrical insulator and shows no obvious signs of electronic transitions. This is as expected for the formation of a condensed valence bond state (which occurs as T = 96 K). The high temperature data was fit to a model $\rho = \rho_o e^{E_g/2K_BT}$ giving an estimated band gap of $E_g$ = 0.12 eV.

**Magnetism and spin-orbit coupling considerations.**

In 4d and 5d systems, spin orbit (SOC) contributions to magnetism can be significant. The spin-orbit coupling constant for Mo is $\xi = 0.068$ eV, similar to the value for Cu, $\xi = 0.100$ eV[37]. Furthermore, the crystal field splitting of adjacent states in a $Mo_3O_{13}$ cluster is ~1.2 eV, on the same scale as in $Cu^{2+}$ compounds (~2.4 eV). Thus the effects of SOC are expected to be similar, within a factor of 2. To more quantitatively assess the expected effect of spin orbit coupling on the observed magnetism, we calculated the expected deviation in the magnetic g-factor arising from a partial unquenched orbital contribution (from SOC) in a perturbative manner. To second order[36], the observed g-value, $g_m$, is given by $g_m = g_e(1 - A\frac{\xi}{\Delta E})$ where $g_e$ is the ideal value (= 2.0), $\xi$ is the spin-orbit coupling constant, and $\Delta E$ is the energy gap between electronic states from crystal field effects, and $A$ is a constant dependant on the exact nature of the ground and low-lying electronic states. The values of $\xi$ are taken from the literature[37] and values for $A$ are typically in the range of 2-4. Performing calculations using the extremes of A, we see that $g_m$ ranges from 1.55 to 1.77, consistent with a g-factor value extracted from the high-temperature Curie constant of 0.24 emu·K·Oe$^{-1}$·mol $f.u.^{-1}$ found in our data ($g_m \approx 1.6$). Performing this same type of calculation on Cu yields a g-factor range of 2.18 to 2.37, consistent with experiment ($g_m \approx 2.2$). This shows that spin-orbit coupling can be an explanation as to the reduced value of the Curie constant (and hence $\mu_{eff}$) observed in our inverse susceptibility data.

However, while spin-orbit coupling can itself give rise to interesting magnetic phases[38], an analysis carried out in the same way as Kotani show that, in this case, it does not explain the observed trends in the magnetic data, notably the transition around $T = 96$ K.

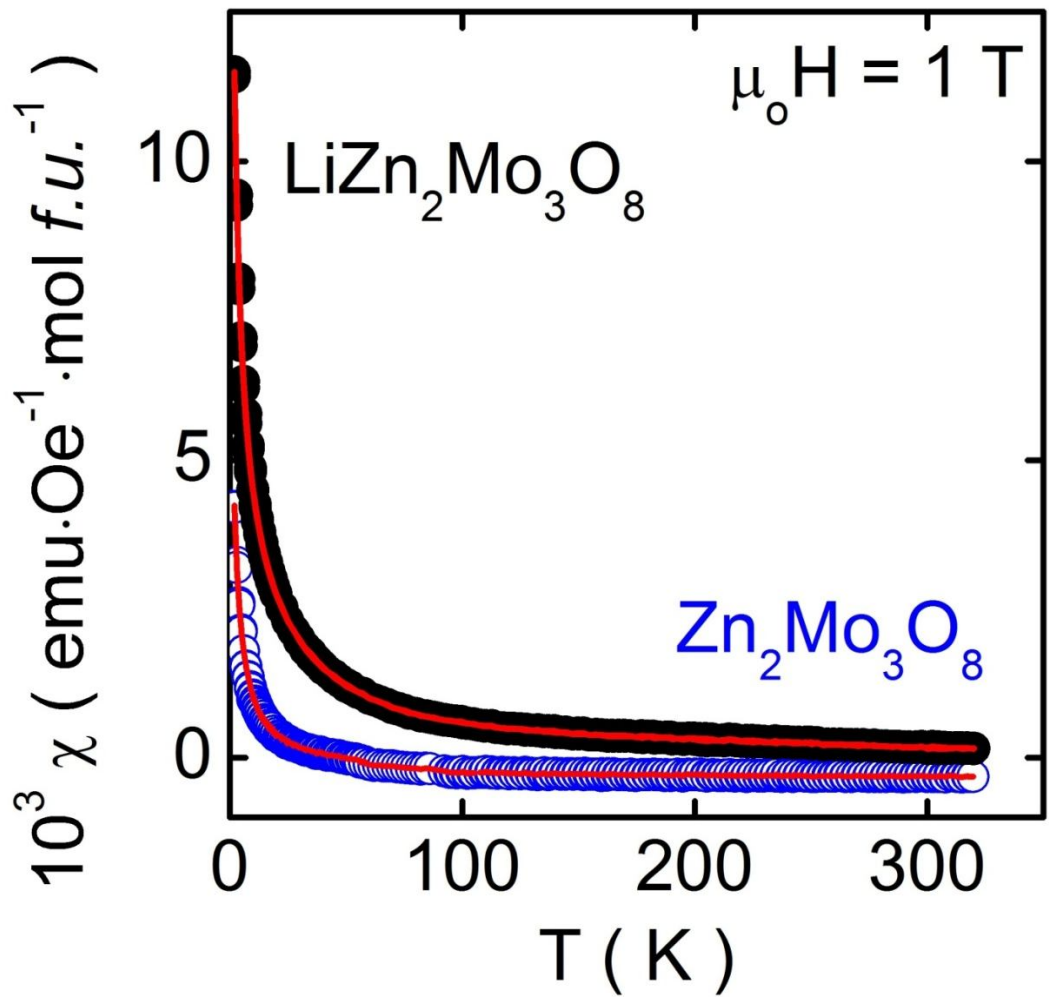

**Figure S7 | LiZn$_2$Mo$_3$O$_8$ and Zn$_2$Mo$_3$O$_8$ magnetic susceptibility.** Magnetic susceptibility data as a function of temperature for LiZn$_2$Mo$_3$O$_8$ and Zn$_2$Mo$_3$O$_8$. DC magnetization was measured on equimolar amounts (0.112 mmol) of each compound using the same sample container. This allowed for direct subtraction of the two datasets to determine the intrinsic response of the unpaired electrons in LiZn$_2$Mo$_3$O$_8$. M(H) curves were linear up to $\mu_oH = 1$ T, the applied field used for the measurements, and thus the susceptibility was calculated assuming $\chi = \dfrac{M}{H}$. The small Curie tail on Zn$_2$Mo$_3$O$_8$ corresponds to 4.4% of paramagnetic impurity spins, mostly from the sample holder.

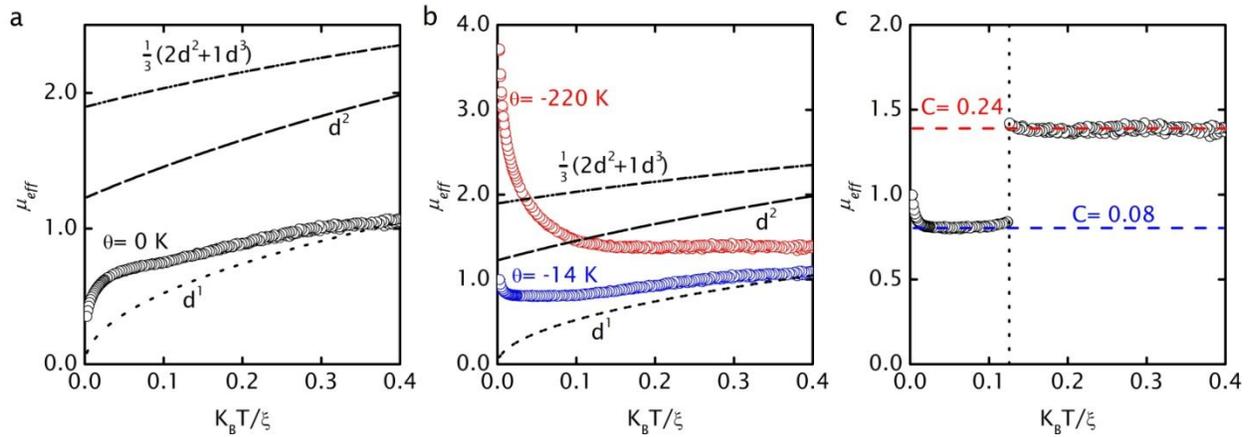

**Figure S8 | LiZn$_2$Mo$_3$O$_8$ temperature dependence of effective magnetic moments.**

Calculations of effective magnetic moments on LiZn$_2$Mo$_3$O$_8$ as a function of thermal energy (k$_B$T) divided by the spin orbit coupling constant ($\xi = 0.068 eV$) for molybdenum. The formula $\mu_{eff} = \sqrt{\frac{3k_B}{N_A \mu_B^2}(\chi - \chi_o)(T - \theta)}$ is used to calculate the effective magnetic moment where $\chi$ is the measured magnetic susceptibility, $\chi_o$ is the temperature independent contribution to the susceptibility (in this case = 0 because of the subtraction of Zn$_2$Mo$_3$O$_8$ and sample holder susceptibility), $T$ is temperature and $\theta$ is the Weiss temperature - indicative of the magnitude and sign of interactions between magnetic moments. $\mu_{eff}$ is in units of $\mu_B$ per f.u. The plots show the calculated magnetic moment for **a,** no interactions ($\theta = 0$ K), **b,** weak ($\theta = -14$ K) or strong ($\theta = -220$K) antiferromagnetic interactions uniform over all temperatures, and **c,** a change from strong to weak antiferromagnetic interactions at $T \approx 96$ K. In **a** and **b,** the expected Kotani behaviors for d$^2$ ions, d$^1$ ions, and a scaled (by one-third) combination of 2 d$^2$ and 1 d$^3$ ions are shown. The d$^1$ case is similar to our proposed spin-½ degree of freedom magnetic molybdenum clusters, and the linear combination of d$^2$ and d$^3$ case is what one would expect if each cluster were comprised of two Mo$^{4+}$ and one Mo$^{3+}$ distinct ions. The observed data are not consistent with any of these Kotani behaviors, as can be seen; the shape of the data is inconsistent with a change in unquenched orbital contribution with temperature (which is the hallmark of SOC Kotani behavior). In **c,** the extracted Weiss temperatures from fits to the inverse susceptibility data are used for the temperature ranges from which they are extracted. In this case the appearance of two flat regions above and below this transition indicates a loss of a large contribution to the effective magnetic moment around this temperature. This supports the argument that $LiZn_2Mo_3O_8$ behaves as an isolated $S=½$ system with a partial unquenched orbital contribution that is temperature-independent (just like $Cu^{2+}$ compounds).

**Table S1 | GAMESS input files:**

GAMESS input files for calculations on a $Mo_3O_4(OH)_3(H_2O)_6$ cluster using unrestricted spin density functional theory with the PBE0 hybrid functional and Popel's N-21 split valence basis set with 3 gaussians for Mo, the 6-31G(d) basis for O and the 6-31G basis for H[39,40]. $Mo_3O_4(OH)_3(H_2O)_6$ was chosen because it is a neutral molecule, has the same electron count as the $Mo_3O_{13}$ clusters in $LiZn_2Mo_3O_8$, and maintains the $C_{3v}$ symmetry. Input file (a) uses Huzinaga's minimal basis set to provide a suitable initial guess for the electron densities, which is then used as input for the full calculation, using input file (b). Changes to the hybrid functional or basis sets used in (b) results in only minor changes to the calculated frontier orbitals. Spin-orbit coupling effects are not included in these calculations.

**a**

```
 $CONTRL SCFTYP=UHF RUNTYP=energy MAXIT=200 ICHARG=0 MULT=2
COORD=UNIQUE UNITS=ANGS EXETYP=RUN DFTTYP=PBE0 $END
 $SYSTEM TIMLIM=525600 MEMORY=10000000 $END
 $BASIS GBASIS=MINI $END
 $GUESS GUESS=HUCKEL $END
 $SCF DIRSCF=.TRUE. FDIFF=.FALSE. SOSCF=.TRUE. DAMP=.TRUE. $END
 $DATA
Mo3O4(OH)3(H2O)6
Cnv 3
0.0 0.0 0.0 0.0 0.0 1.0
1.0 0.0 0.0 'PARALLEL'
Mo   42.0 1.48725 0.00000 0.00000
O4    8.0 0.00000 0.00000 -1.45412
O2    8.0 -1.61790 0.00000 1.2638
O3    8.0 3.35505 0.00000 1.0374
O1    8.0 2.57769 1.3475 -1.11802
H3    1.0 3.35505 0.00000 2.00706
H1_1  1.0 2.18474 1.17574 -1.98769
H1_2  1.0 2.66704 2.21202 -0.68806
 $END
```

**b**

```
 $CONTRL SCFTYP=UHF RUNTYP=energy MAXIT=200 ICHARG=0 MULT=2
COORD=UNIQUE UNITS=ANGS EXETYP=RUN DFTTYP=PBE0 ISPHER=1 $END
 $SYSTEM TIMLIM=525600 MEMORY=10000000 $END
 $BASIS BASNAM(1)=metal,metal,metal,ligO,
                  ligO,ligO,ligO,
                  ligO,ligO,ligO,
                  ligO,ligO,ligO,ligO,ligO,ligO,
                  ligH,ligH,ligH,
                  ligH,ligH,ligH,ligH,ligH,ligH,
                  ligH,ligH,ligH,ligH,ligH,ligH $END
 $GUESS GUESS=RDMINI $END
 $SCF DIRSCF=.TRUE. FDIFF=.FALSE. DIIS=.TRUE. DAMP=.TRUE. $END
 $DATA
Mo3O4(OH)3(H2O)6
Cnv 3
0.0 0.0 0.0 0.0 0.0 1.0
1.0 0.0 0.0 'PARALLEL'
Mo   42.0 1.48725 0.00000 0.00000
O4    8.0 0.00000 0.00000 -1.45412
O2    8.0 -1.61790 0.00000 1.2638
O3    8.0 3.35505 0.00000 1.0374
O1    8.0 2.57769 1.3475 -1.11802
H3    1.0 3.35505 0.00000 2.00706
H1_1  1.0 2.18474 1.17574 -1.98769
H1_2  1.0 2.66704 2.21202 -0.68806
 $END
 $metal
n21 3

 $end
 $ligO
n31 6
d 1 ; 1 0.8 1.0

 $end
 $ligH
n31 6

 $end

 $VEC
```

**Table S2 | Crystallographic information**

| LiZn$_2$Mo$_3$O$_8$ refinement | |
|---|---|
| Chemical formula sum | Li1.2(1) Zn1.8(1) Mo3 O8 |
| Space group | $R\bar{3}m$ |
| a (Å) | 5.7956(3) |
| b (Å) | 5.7956(3) |
| c (Å) | 31.039(3) |
| Z | 6 |
| wR$_p$ | 0.0303 |
| R$_p$ | 0.0445 |
| R(F$^2$) | 0.0846 |
| $\chi^2$ | 8.356 |
| LeBail $\chi^2$ | 2.705 |

| LiZn$_2$Mo$_3$O$_8$ structural parameters | | | | | | |
|---|---|---|---|---|---|---|
| atom | x | y | z | Wyck. pos. | Occ | U$_{iso}$ |
| Mo | 0.18573(7) | 0.81428(7) | 0.08401(4) | 18h | 1.000 | 0.0033(1) |
| O1 | 0.84500(9) | 0.15498(9) | 0.04766(3) | 18h | 1.000 | 0.0012(1) |
| O2 | 0.49217(10) | 0.50783(10) | 0.12404(3) | 18h | 1.000 | 0.0060(1) |
| O3 | 0.0000 | 0.0000 | 0.11820(7) | 6c | 1.000 | 0.0052(2) |
| O4 | 0.0000 | 0.0000 | 0.37178(6) | 6c | 1.000 | 0.0029(2) |
| Zn1 | 0.3333 | 0.6667 | -0.64148(6) | 6c | 0.901(4) | 0.0014(1) |
| Zn2 | 0.0000 | 0.0000 | 0.18132(9) | 6c | 0.716(4) | 0.0014(1) |
| Li1 | 0.3333 | 0.6667 | -0.64148(6) | 6c | 0.099(4) | 0.0014(1) |
| Zn3 | 0.0000 | 0.0000 | 0.0000 | 3a | 0.226(7) | 0.0014(1) |
| Li2 | 0.0000 | 0.0000 | 0.18132(9) | 6c | 0.284(4) | 0.0014(1) |
| Zn4 | 0.0000 | 0.0000 | 0.5000 | 3a | 0.143(6) | 0.0014(1) |
| Li3 | 0.0000 | 0.0000 | 0.5000 | 3a | 0.774(7) | 0.0014(1) |
| Li4 | 0.0000 | 0.0000 | 0.5000 | 3a | 0.857(6) | 0.0014(1) |

# Supplementary References